\documentclass{jpsj3}
\usepackage{txfonts}
\usepackage{bm}
\usepackage{graphicx}
\usepackage{color}
\usepackage{float}

\title{Kernel Learning by Quantum Annealer}

\author{Yasushi Hasegawa$^{1,2}$\thanks{hasegawa.yasushi.s6@dc.tohoku.ac.jp}, Hiroki Oshiyama$^1$, and Masayuki Ohzeki$^{1,3,4}$}
\inst{$^1$Graduate School of Information Sciences, Tohoku University, Sendai 980-8579, Japan \\
$^2$Technology Strategy Division, SCSK Corporation, Koto, Tokyo 135-8110, Japan \\
$^3$Department of Physics, Institute of Science Tokyo, Meguro, Tokyo 152-8551, Japan \\
$^4$Sigma-i Co., Ltd., Minato, Tokyo 108-0075, Japan} 

\abst{The Boltzmann machine is one of the various applications using a quantum annealer. As a feasibility study, we propose an application of the Boltzmann machine to the kernel matrix used in various machine learning techniques. We focus on the fact that shift-invariant kernel functions can be expressed in terms of the expected value of a spectral distribution by the Fourier transformation. Using this transformation, the random Fourier feature (RFF) samples the frequencies and approximates the kernel function. Furthermore, this paper proposes a method to obtain a spectral distribution suitable for the data using a Boltzmann machine. Across Fashion MNIST binary tasks, our approach achieves top or comparable accuracy to a tuned Gaussian RFF and Random Kitchen Sinks with Implicit Kernel Learning, while consistently attaining lower loss and learning bimodal, data-adaptive spectral distributions. On synthetic data (d=10), it improves test accuracy by +2.8 points over Gaussian RFF. These results indicate that our method remains valuable even when accuracy saturates, by yielding better kernel alignment and richer spectral structure, and it is practically enabled by fast sampling on a quantum annealer.}

\begin{document}
\maketitle

\section{Introduction}

Quantum annealing is a heuristic algorithm that searches for the ground state of a predetermined Hamiltonian by using quantum tunneling effects \cite{Ray1989,Kadowaki1998,Ohzeki2015}. It has been used in numerous applications\cite{Yarkoni2022}, including portfolio optimization\cite{Rosenberg2016,Venturelli2019}, molecular similarity problem\cite{Hernandez2017}, quantum chemical calculation\cite{Streif2019}, preprocessing in material experiments\cite{Tanaka2023}, scheduling problem\cite{Venturelli2016,Ikeda2019,Yarkoni2021}, traffic optimization\cite{Stollenwerk2020,Inoue2021}, machine learning\cite{Neven2009,neven2009nips,Neukart2018,Crawford2019,Sato2021}, web recommendation\cite{Nishimura2019}, steel manufacturing \cite{Yonaga2022}, and route optimization for automated guided vehicles in factories\cite{Ohzeki2019,Haba2022} as well as in decoding problems \cite{Ide2020,Arai2021cdma}. A comparative study of the quantum annealer with other solvers was performed for benchmark tests to solve optimization problems \cite{Oshiyama2022}. The quantum effect on the case with multiple optimal solutions has also been discussed \cite{Yamamoto2020, Maruyama2021}. As the environmental effect cannot be avoided, the quantum annealer is sometimes regarded as a simulator for quantum many-body dynamics \cite{Bando2020,Bando2021,King2022}.

It is known that the output from the quantum annealer follows a Gibbs-Boltzmann distribution due to noise from thermal effects and residual magnetic fields from freezing effects. For this reason, quantum annealing may generate samples that follow the Gibbs-Boltzmann distribution instead of searching for an optimal solution.
Among applications of quantum annealing for machine learning for solving optimization problems have been reported, a typical application that uses the Gibbs-Boltzmann distribution is the Boltzmann machines\cite{Crawford2019,Adachi2015,benedetti2016estimation, Arai2021,Sato2021,Urushibata2022}. 
The Boltzmann machine is a learning method that approximates the Gibbs-Boltzmann distribution prepared as a model to the empirical distribution of the given dataset.
In particular,a restricted Boltzmann machine (RBM) is a Boltzmann machine in which the nodes are divided into visible and hidden layers and the connections of the nodes are restricted within each layer). 

The RBM is a widely used machine learning model for unsupervised and supervised tasks\cite{Adachi2015,Salakhutdinov2007,Larochelle2008}. 
In RBM, the loss function uses the Kullback-Leibler divergence between the dataset's frequency distribution and the Gibbs-Boltzmann distribution. The model is updated based on the difference between the dataset's expected values and the expected values given by the Gibbs-Boltzmann distribution. However, calculating the expected value of the Gibbs-Boltzmann distribution requires calculating the frequency of the states provided by the Gibbs-Boltzmann distribution, so it is generally computationally expensive.
It takes advantage of the fact that sampling from a quantum annealer follows a Gibbs-Boltzmann distribution, which allows for faster learning by approximating the model-dependent term as the expected value of the sample. Therefore, by utilizing the fact that sampling from a quantum annealer follows the Gibbs-Boltzmann distribution, it is possible to approximate the expected values given by the Gibbs-Boltzmann distribution with samples obtained from quantum annealing, which enables faster learning.

This paper proposes a new application focusing on the output distribution of a quantum annealer: kernel learning. Kernel methods are powerful tools in machine learning. 
Although selecting a kernel function that fits the data is crucial, there is no systematic approach to choosing the best one. Therefore, kernel learning was proposed to learn a kernel function that fits the data \cite {Li2019,Mairal2016,Gonen2011}. 
We employ a multi-layer RBM as the model to learn the kernel function and use a quantum annealer to train the RBM. In our experiment, we compared our method with a kernel based on a Gaussian distribution and with Random Kitchen Sinks (RKS) with Implicit Kernel Learning (IKL)\cite{Li2019}, as Gaussian kernels are widely used as a standard choice in kernel methods. The classification accuracy of our approach was comparable to that of a parameter-tuned Gaussian kernel and RKS with IKL; however, our method offers broader applicability as it can adapt to the given data. Furthermore, unlike fault-tolerant quantum computers, quantum annealers provide a practical and near-term alternative for exploring quantum-enhanced machine learning.

The remainder of this paper is organized as follows: In the Methods section, we formulate kernel learning using multi-layer RBM and present necessary tools such as kernel methods, random Fourier feature (RFF), kernel learning, and quantum annealing. In the following section, we demonstrate our method for the binary classification task on synthetic data and the Fashion MNIST dataset and compare its accuracy with the RFF model using a Gaussian distribution and with RKS with IKL.
The last section summarizes our study and discusses potential future work.

\section{Methods}

This section describes the details of kernel learning using the restricted Boltzmann machine (RBM). The key novelty of our work lies in leveraging the Gibbs-Boltzmann distribution generated by quantum annealers to learn spectral distributions adaptively. We construct data-adaptive kernel functions using these distributions via random Fourier feature (RFF). This methodology replaces traditional fixed kernel functions with learned ones tailored to the specific data. We briefly describe several ingredients to achieve this goal, such as kernel methods, RFF, kernel learning, and quantum annealing.

\subsection{Kernel Methods}

\begin{figure}[H]
\includegraphics[width=\linewidth]{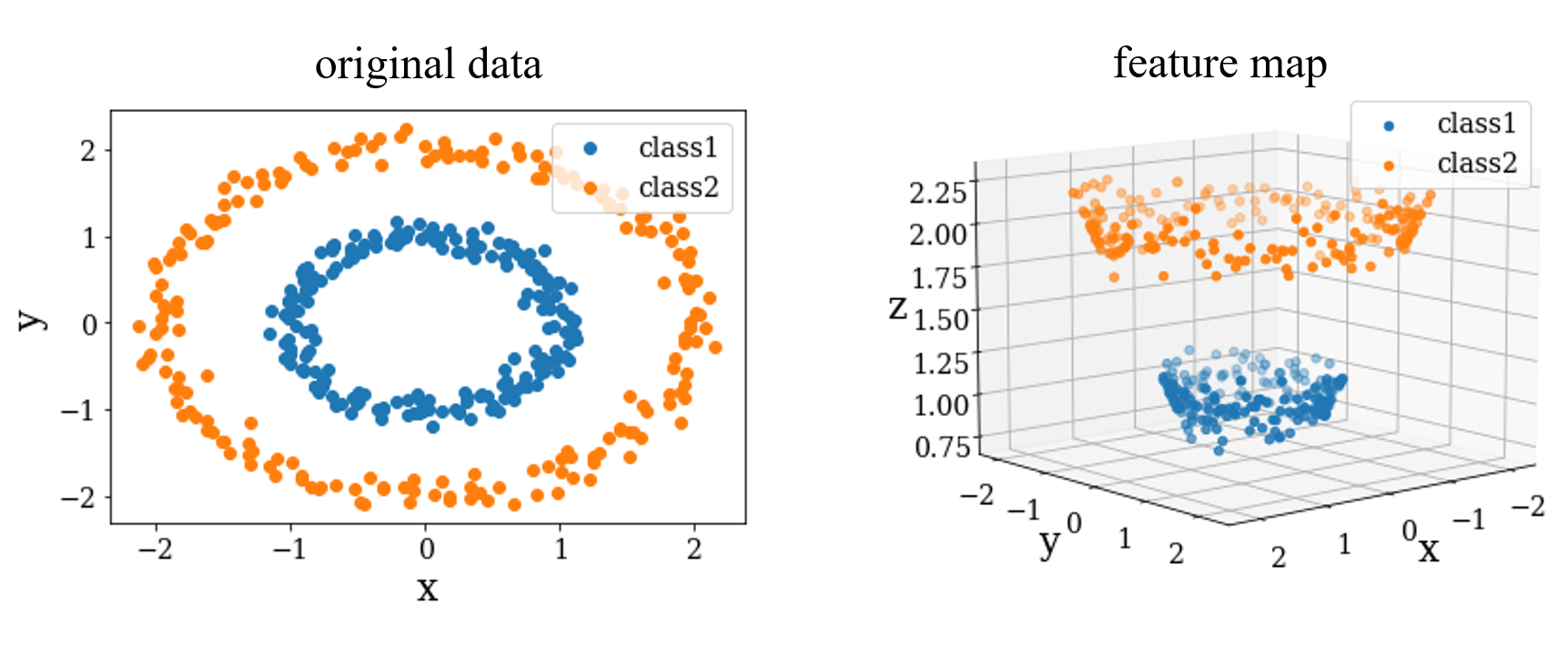}
\caption{(Color online) Mapping to feature space by the kernel function. The left panel represents a two-dimensional data set that is not linearly separable, and the right panel shows an example of how it can be transformed into a three-dimensional data set with the addition of $ z=\sqrt{x^{2}+y^{2}}$ dimension to make it linearly separable.}
\label{fig:feature map}
\end{figure}

The perceptron is a machine learning model that inputs multiple signals and outputs a single signal.
Using the weight vector $ \bm{W} $ for the signal, it can be expressed as $ f(\bm{x}) = \text{sign}( \bm{Wx} )$. In perceptron training, the weight $ \bm{W} $ is updated so that the correct label $ y $ can be output for a given data $ \bm{x} $. 
Here, the weight $ \bm{W} $ after learning is $ \bm{W} = \sum_{i}^{N} {\alpha_{i} y_{i} \bm{x}_{i} } $ with real vector $ \bm{\alpha} $. 
As a result, the perceptron can be written as $ f( \bm{x} ) = \text{sign} ( \sum_{i}^{N} {\alpha_{i} y_{i} \bm{x}_{i}} \bm{x})$. Since a linear combination of input signals represents the perceptron, it can only solve linearly separable problems. 
To deal with linearly inseparable problems, we introduce a nonlinear function $ \phi $ and consider $ F( \bm{x} ) = \text{sign} ( \bm{U}\phi(\bm{x} ))$. By following the same procedure as $ f( \bm{x} ) $, $ F( \bm{x} ) = \text{sign} ( \sum_{i}^{N} { \alpha_{i} y_{i} \phi( \bm{x}_{i} )} \phi( \bm{x} ))$. Here, $ F( \bm{x} ) $ can be determined from the inner product $ \phi( \bm{x}_{i} ) \phi( \bm{x} ) $ of the data points. Therefore, instead of a nonlinear function, it can be replaced by a kernel function $k( \bm{x},\bm{x'} )$ defined from two data points. The kernel function is a function that can be defined arbitrarily if it has the property of being an inner product. The well-known kernel functions are linear, sigmoidal, polynomial, and Gaussian kernels\cite{Karal2020}. 
Especially for classification tasks, it is crucial to collect the same data points and map them to a linearly separable space as shown in Figure \ref{fig:feature map}. 

\subsection{Random Fourier Feature}

Random Fourier feature is the method that explicitly defines the feature map $ \varphi $ and approximates the kernel as $ k( \bm{x}, \bm{x'} ) \approx \varphi( \bm{x}) \varphi( \bm{x'}) $. In general, the kernel function is computed for a combination of two data points $ \bm{x} $ and $ \bm{x'} $; consequently, its computational complexity is $ O(N^{2} ) $ for N data points. On the other hand, if the dimension of $ \varphi( \bm{x} ) $ is $ S $ when using RFF and $ S $ is smaller than $ N $, its computational complexity can be reduced to $ O(NS) $\cite{Liu2020}.

In this method, we focus on the shift-invariant kernel functions $ k(\bm{x}, \bm{x'})=k(\bm{x} - \bm{x'}) $ among kernel functions. 
The shift-invariant kernel is expressed as follows, using the expected value of the spectral distribution from Bochner's theorem \cite{bochner1959}.
\begin{equation}
k(\bm{x} - \bm{x'}) = \int_{\mathbb{R}_{d}} e^{i\bm{\omega}(\bm{x} - \bm{x'})} dp(\bm{\omega}) = E_{\bm{\omega} \sim p(\bm{\omega})}[e^{i{\bf \omega}(\bm{x} - \bm{x'})}],
\end{equation}
where $ \bm{\omega} $ is a random vector of the same dimension as $ \bm{x} $ sampled from the spectral distribution $ p (\bm{\omega}) $. 
By using a finite set $\{ \bm{\omega} \}$, a low-variance approximation of the above equation can be obtained \cite{Rahimi}.
\begin{equation}
k(\bm{x} - \bm{x'}) = E_{\bm{\omega} \sim p(\bm{\omega})}[e^{i \bm{\omega}(\bm{x} - \bm{x'}))}] \approx  \frac{1}{S} \sum_{s}e^{i \bm{\omega}_{s}(\bm{x} - \bm{x'})}.
\end{equation}
Furthermore, since the kernel function is used as a real-valued function, only the real-valued part of the above equation is employed. Thus,
\begin{equation}
\left[\frac{1}{S} \sum_{s} e^{i \bm{\omega}_{s}(\bm{x} - \bm{x'})}\right]_\mathbb{R}=\frac{1}{S} 
\sum_{s}\cos\bm{\omega}_{s}(\bm{x} - \bm{x'}) = \varphi (\bm{x}) \varphi (\bm{x'}).
\end{equation}
The function $ \varphi (\bm{x}) $ is defined as follows.
\begin{equation}
\varphi(\bm{x})=\frac{1}{\sqrt{S}}(\cos\bm{\omega}_{1}\bm{x}, \cos\bm{\omega}_{2}\bm{x}, \cdots, \cos\bm{\omega}_{s}\bm{x}, \sin\bm{\omega}_{1}\bm{x}, \sin\bm{\omega}_{2}\bm{x}, \cdots, \sin\bm{\omega}_{s}\bm{x}).
\end{equation}
As a result, the shift-invariant kernel is represented by the feature map $ \varphi$ that can be derived by using a random finite set $ \{ \omega \} $, and its computational complexity can be set to $ O(NS) $.

\subsection{Kernel Learning}

We describe kernel learning that utilizes the spectral distribution $ p (\bm{\omega}) $ to obtain a kernel function suitable for the data. A kernel function can be defined arbitrarily as long as it is expressed in the inner product. On the other hand, there is no systematic approach to selecting the best kernel function that fits the data. Instead of heuristically selecting the kernel function, Multiple Kernel Learning combines existing kernels to obtain a better kernel function that fits the data \cite{Gonen2011}. 

Recently, there is also a method called Implicit Kernel Learning (IKL), which learns a kernel function by learning the spectral distribution of the kernel function\cite{Li2019}. Implicit Kernel Learning is a new method that models spectral distributions by learning the sampling process. 
Thus, kernel selection can be replaced by optimized learning of the spectral distribution.
\begin{equation}
\arg \max_{k \in K} \sum_{i=1} E_{\bm{x} \sim p_{i}, \bm{x'} \sim q_{i}} [F_{i}(\bm{x}, \bm{x'}) k(\bm{x}, \bm{x'})] = \arg \max_{k \in K} \sum_{i=1} E_{\bm{x} \sim p_{i}, \bm{x'} \sim q_{i}} \left[F_{i}(\bm{x}, \bm{x'}) E_{\bm{\omega} \sim p(\bm{\omega})}[e^{i \bm{\omega} (\bm{x} - \bm{x'})}] \right],
\end{equation}
where function $ F_{i}(\bm{x}, \bm{x'}) $ represents the learning task-specific objective function\cite{Li2019}. To perform the classification task in this study, we set $ F_{i} (\bm{x}, \bm{x'}) = yy' $, where $ y $ and $ y' $ are the labels $\{ +1, -1 \}$ of the data $ \bm{x} $ and $ \bm{x'} $, respectively. This corresponds to treating $ \bm{x}, \bm{x'} $ as similar data in the feature space by maximizing the above equation when $ y, y' $ have the same label and treating $ \bm{x}, \bm{x'} $ as different data in the feature space by minimizing the above equation when $ y, y' $ have different labels. 

Our method consists of two steps. In step 1, kernel learning is performed through updates of the spectral distribution $ p(\bm{\omega}) $, and in step 2, classification learning is performed using the learned kernel function.

\subsection{Quantum Annealing}

In step 1, quantum annealing is used to learn the kernel function. In general, quantum annealing is an algorithm that searches for the ground state of the Ising model shown below by using quantum effects\cite {Ray1989}.
\begin{equation}
H_{\text{ising}} = \sum_{i} h_{i} {\hat{\sigma}}_z^{i} + \sum_{ij(>i)} J_{ij} {\hat{\sigma}}_{z}^{i} {\hat{\sigma}}_{z}^{j},
\end{equation}
where $ \hat{\sigma}_{z}^{i} $ is the Pauli matrix on the i-th qubit, $ h_{i} $ and $ J_{ij} $ are the bias on each qubit and coupling strength between each pair of qubits, respectively. In actual quantum annealers, the following Hamiltonian, which adds $ \sum_{i} {\hat{\sigma}}_x^{i} $ to the Ising model, is used.
\begin{equation}
H = \frac{A(s)}{2} \sum_{i}{\hat{\sigma}}_{x}^{i} + \frac{B(s)}{2} \left( \sum_{i} h_{i} {\hat{\sigma}}_z^{i} + \sum_{ij(>i)} J_{ij} {\hat{\sigma}}_{z}^{i} {\hat{\sigma}}_{z}^{j} \right),
\end{equation}
where $ \hat{\sigma}_{x}^{i} $ is the Pauli matrix on the i-th qubit, and $s$ is called the annealing rate. $ A(s) $ and $ B(s) $ are known as annealing functions. At the initial state of quantum annealing, i.e., when $ s = 0 $, the annealing function is $ A(s) \gg B(s) $, and the qubits are in a global superposition state that is the ground state of $ \sum_{i} {\hat{\sigma}}_x^{i} $. Then, as $ s $ is increased from $ 0 $ to $ 1 $, the ground state of the Hamiltonian changes. Finally, when $ s = 1 $, $ A(s) \ll B(s) $, the qubits converge to a single classical state. At this point, the state $ \hat{\sigma}_{z}^{i} $ of the Ising model is obtained.

However, the output of an actual quantum annealer is not necessarily the ground state of the above Ising model.
The distribution of the outputs instead follows the Gibbs-Boltzmann distribution due to thermal effects and residual magnetic fields due to the freezing effect \cite{amin2015searching}.

\subsection{Kernel Learning with Multi-layer RBM}

\begin{figure}[H]
\centering
\includegraphics[width=\linewidth]{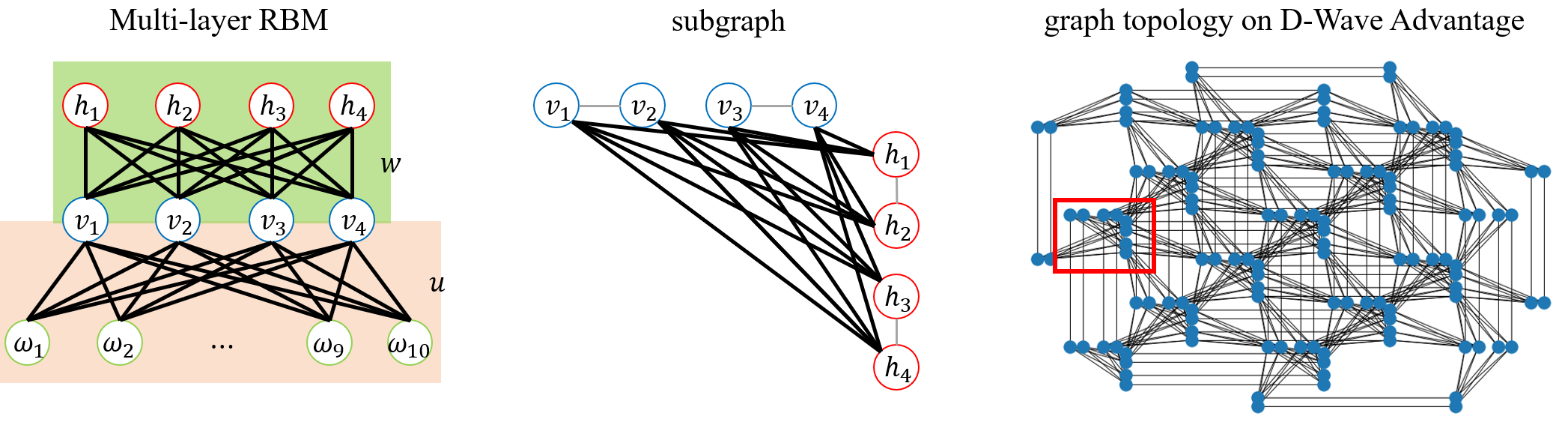}
\caption{(Color online) Boltzmann machine and minor embedding. The figure on the left panel is a multi-layer RBM. The number of nodes is $\{ N_{\bm{\omega}}, N_{\bm{v}}, N_{\bm{h}} \} = \{ 10, 4, 4 \}$. $ w $ and $ u $ are matrices of coupling strength between $ \bm{v} $, $ \bm{h} $ and $ \bm{\omega} $, $ \bm{v} $ respectively. Right panel shows the graph topology on D-Wave Advantage and subgraph. The bold lines represent the interaction. In the experiment, the network of $ \bm{v} $, $ \bm{h} $ in the left panel is embedded in the highlighted subgraph in the right panel, as shown in the middle panel. The output $ \bm{v} $ is then used to perform $ \bm{\omega} $ sampling on a classical computer according to equation (10).}
\label{fig:boltzmann machine}
\end{figure}

In this study, to convert the samples obtained from the quantum annealer to $ \bm{\omega} $, we use the multi-layer RBM as shown in the left-hand side of Figure \ref{fig:boltzmann machine}. The Gibbs-Boltzmann distribution that the output of a quantum annealer follows is expressed as follows.
\begin{equation}
P(\bm{v}, \bm{h}| \bm{b}, \bm{c}, w) = \frac{e^{-E(\bm{v}, \bm{h}| \bm{b}, \bm{c}, w)}}{Z (\bm{b}, \bm{c}, w)}
\end{equation}
with
\begin{equation}
E(\bm{v}, \bm{h}| \bm{b}, \bm{c}, w) = - \sum_{i}{b_{i} v_{i}} - \sum_{j}{c_{j} h_{j}} - \sum_{ij}{v_{i}w_{ij}h_{j}},
\end{equation}
where $ Z $ is the partition function, $ b_{i}, c_{j} $ are bias terms, and $ w_{jk} $ is the coupling strength between variables $ v_{i} $ and $ h_{j} $. After sampling from the Gibbs-Boltzmann distribution, a Gaussian-Bernoulli type RBM (GB-RBM)\cite{Cho2011,choo2018,abdullah2022} is used to transform the discrete variable $ \bm{v} $ into a continuous variable $ \bm{\omega} $ on a classical computer using the following equation.
\begin{equation}
P(\bm{\omega}| \bm{a}, u, \bm{v}, \bm{\sigma}) = \frac{e^{-E(\bm{\omega}| \bm{a}, u, \bm{v}, \bm{\sigma})}}{Z (\bm{a}, u, \bm{v}, \bm{\sigma})} = \frac{1}{\sqrt{2\pi} \bm{\sigma}} e^{- \frac{\left(\bm{\omega} - (\bm{a} + U\bm{v})\right)^{2}}{2 \bm{\sigma}^{2}}}
\end{equation}
with
\begin{equation}
E(\bm{\omega}| \bm{a}, \bm{v}, u, \bm{\sigma}) =  \sum_{i} \frac{(\omega_{i} - (a_{i} + \sum_{j}u_{ij}v_{j}))^{2}}{2\sigma_{i}^{2}},
\end{equation}
where $ a_{i} $ is the bias and $ u_{ij} $ is the coupling strength between variables $ \omega_{i} $ and $ v_{j} $, respectively. $ \sigma_{i} $ is the standard deviation with respect to variable $ \omega_{i} $. Specifically, the number of nodes in the RBM was set to $ \{ N_{\bm{\omega}}, N_{\bm{v}}, N_{\bm{h}} \} = \{ 10, 4, 4 \} $, as shown in the left-hand side of Figure \ref{fig:boltzmann machine}. D-Wave Advantage is used for sampling the visible layers $\{ v_{1}, v_{2}, v_{3}, v_{4} \}$ and the hidden layers $\{ h_{1}, h_{2}, h_{3}, h_{4} \}$. 

The D-Wave Advantage has a graph structure where the qubits are not fully connected, as shown on the right-hand side of Figure \ref{fig:boltzmann machine}. Since there are subgraphs in the graph that can form a bipartite graph with eight qubits, the visible and hidden layers were embedded in that bipartite graph. When performing quantum annealing, the Hamiltonian was constructed with $ \{ h_{1}, \cdots , h_{8} \} = 
\{ b_{1}, \cdots, b_{4}, c_{1}, \cdots, c_{4} \} $ and $ \{ J_{14}, \cdots, J_{48} \} = \{ w_{11}, \cdots, w_{44} \} $. As a result of the sampling, $ ( \hat{\sigma}_{z}^{1}, \cdots, \hat{\sigma}_{z}^{8} ) = \{ v_{1},\cdots, v_{4}, h_{1}, \cdots, h_{4} \} $ are obtained.

Next, we discuss the RBM parameter updates. 
In this experiment, the loss function is defined in the section Kernel Learning.
\begin{equation}
L = \frac{1}{N^{2}} \sum_{ij} {y_{i} y_{j} E_{\bm{\omega}} [e^{i{\bf \omega}(\bm{x}_{i} - \bm{x}_{j})}]}.
\end{equation}
The gradient of the loss function is expressed below using the Boltzmann machine parameter $ \theta$.
\begin{equation}
\frac{\partial L}{\partial \theta} = \frac{1}{N^{2}} \sum_{ij} {y_{i} y_{j} \frac{\partial}{\partial \theta} E_{\bm{\omega}} [e^{i{\bf \omega}(\bm{x}_{i} - \bm{x}_{j})}]}.
\end{equation}
Here, approximating the expectation calculation by sampling $\{ \omega_{1}, \omega_{2}, \cdots, \omega_{S} \}$, the gradient can be calculated as follows,
\begin{equation}
\frac{\partial L}{\partial \theta} = \frac{1}{N^{2}} \sum_{ij} {y_{i} y_{j} \frac{\partial}{\partial \theta} E_{\bm{\omega}} [e^{i{\bf \omega}(\bm{x}_{i} - \bm{x}_{j})}]} = \frac{1}{N^{2}} \sum_{ij} {\sum_{s} {y_{i} y_{j} e^{i{\bf \omega_{s}}(\bm{x}_{i} - \bm{x}_{j})} \frac{\partial}{\partial \theta} P(\omega_{s}| \theta)}}.
\end{equation}

In this study of RBM learning, the probability distribution is expressed as follows,
\begin{equation}
P(\bm{\omega}| \bm{a}, u, \bm{v}, \bm{\sigma}) P(\bm{v}, \bm{h}| \bm{b}, \bm{c}, w) = \frac{e^{-E(\bm{\omega}| \bm{a}, \bm{v}, u, \bm{\sigma}) - E(\bm{v}, \bm{h} |\bm{b}, \bm{c}, w)}}{\sqrt{2 \pi} \bm{\sigma} Z(\bm{b}, \bm{c}, w)} = P(\bm{\omega}, \bm{v}, \bm{h}| \bm{a}, \bm{b}, \bm{c}, w, u, \bm{\sigma})
\end{equation}
with
\begin{equation}
E(\bm{\omega}, \bm{v}, \bm{h} | \bm{a}, \bm{b}, \bm{c}, w, u, \bm{\sigma}) = \sum_{i} \frac{(\omega_{i} - (a_{i} + \sum_{j}u_{ij}v_{j}))^{2}}{2\sigma_{i}^{2}} - \sum_{j} {b_{j} v_{j}} - \sum_{k} {c_{k} h_{k}} - \sum_{jk} {v_{j} w_{jk} h_{k}}.
\end{equation}
From $ \sigma_{i}^{2} \geq 0 $, we parametrize $ \sigma_{i}^{2}$ as $ e^{z_{i}}$ and write $ \{ \bm{a}, \bm{b}, \bm{c}, w, u, \bm{z} \} = \theta $.

The gradient $ \frac{\partial}{\partial \theta} P( \bm{\omega}, \bm{v}, \bm{h}| \theta) = \frac{\partial}{\partial \theta}  e^{-E( \bm{\omega}, \bm{v}, \bm{h}| \theta )} / Z(\theta)$ of the Gibbs-Boltzmann distribution can then be calculated as follows,
\begin{equation}
\frac{\partial}{\partial \theta} \frac{e^{-E(\bm{\omega}, \bm{v}, \bm{h}| \theta)}}{Z(\theta)} = \left\{- \frac{\partial}{\partial \theta} E(\bm{\omega}, \bm{v}, \bm{h}| \theta) - \frac{1}{Z(\theta)} \frac{\partial}{\partial \theta} Z(\theta)\right\} \frac{e^{- E(\bm{\omega}, \bm{v}, \bm{h}| \theta)}}{Z(\theta)} = \left\{ - \frac{\partial}{\partial \theta} E(\bm{\omega}, \bm{v}, \bm{h}| \theta) - \frac{1}{Z(\theta)} \frac{\partial}{\partial \theta} Z(\theta)\right\} p(\bm{\omega}, \bm{v}, \bm{h}| \theta).
\end{equation}
Thus, the gradient of the loss function is as follows. The calculation of each parameter is described in the appendix.
\begin{equation}
\frac{\partial L}{\partial \theta} = \frac{1}{N^{2}} \sum_{ij}{\frac{1}{S}\sum_{s}{y_{i} y_{j} e^{i\bm{\omega}_{s} (\bm{x}_{i} - \bm{x}_{j})}}} \left\{-\frac{\partial}{\partial \theta} E(\bm{\omega}, \bm{v}, \bm{h}| \theta) - \frac{1}{Z(\theta)} \frac{\partial}{\partial \theta} Z(\theta)\right\}_{\bm{\omega}_{s}}.
\end{equation}
Similarly to equation (3), the gradient of the loss function can also be separated for the indices $ i $ and $ j $.
\begin{equation}
\frac{\partial L}{\partial \theta} = \frac{1}{N^{2}} \sum_{ij}{\frac{1}{S}\sum_{s}{y_{i} y_{j} e^{i\bm{\omega}_{s} (\bm{x}_{i} - \bm{x}_{j})}}} \left\{-\frac{\partial}{\partial \theta} E(\bm{\omega}, \bm{v}, \bm{h}| \theta) - \frac{1}{Z(\theta)} \frac{\partial}{\partial \theta} Z(\theta)\right\}_{\bm{\omega}_{s}} = \frac{1}{N^{2}} \sum_{i} {y_{i} \psi(\bm{x}_{i}) \sum_{j} {y_{j} \varphi(\bm{x}_{j})}},
\end{equation}
where $ \psi( \bm{x} )$ is defined;
\footnotesize
\begin{multline}
\psi(\bm{x}) = \frac{1}{\sqrt{S}} \Biggl(\left\{ - \frac{\partial}{\partial \theta} E(\bm{\omega}, \bm{v}, \bm{h}| \theta) - \frac{1}{Z(\theta)} \frac{\partial}{\partial \theta} Z(\theta)\right\}_{\bm{\omega}_{1}} \cos{\bm{\omega}_{1} \bm{x}}, \cdots ,\left\{- \frac{\partial}{\partial \theta} E(\bm{\omega}, \bm{v}, \bm{h}| \theta) - \frac{1}{Z(\theta)} \frac{\partial}{\partial \theta} Z(\theta)\right\}_{\bm{\omega}_{S}} \cos{\bm{\omega}_{S} \bm{x}}, \\
\left\{ - \frac{\partial}{\partial \theta} E(\bm{\omega}, \bm{v}, \bm{h}| \theta) - \frac{1}{Z(\theta)} \frac{\partial}{\partial \theta} Z(\theta)\right\}_{\bm{\omega}_{1}} \sin{\bm{\omega}_{1} \bm{x}}, \cdots ,\left\{- \frac{\partial}{\partial \theta} E(\bm{\omega}, \bm{v}, \bm{h}| \theta) - \frac{1}{Z(\theta)} \frac{\partial}{\partial \theta} Z(\theta)\right\}_{\bm{\omega}_{S}} \sin{\bm{\omega}_{S} \bm{x}} \Biggl).
\end{multline}
\normalsize

In step 2, classification learning is performed using the learned kernel function. The classifier uses a kernel perceptron, which is a kernelized perceptron. 

The given data points are transformed with the feature function $ \phi(\bm{x}) $ acquired by kernel learning, and the inner product with the weight matrix $ \sum_{i}^{N} { \bm{\alpha}_{i} y_{i} \phi(x_{i})} $ is calculated. Initially, $ \bm{\alpha} = ( \alpha_{1}, \alpha_{2}, \cdots, \alpha_{N} )$ is set as an all-zero vector, and the value is updated when the output is different from the label. 
This operation is repeated for the number of iterations.
\begin{equation}
f(\bm{x}) = sign\left(\sum_{i}^{N} \alpha_{i} y_{i} \phi(\bm{x}_{i}) \phi(\bm{x})\right) = sign\left(\sum_{i}^{N}\alpha_{i} y_{i} k(\bm{x}_{i},\bm{x})\right).
\end{equation}
If $ f(\bm{x}) $ is different from the label $ y $, the parameter is updated by adding the learning rate to $ \alpha $ and using the new $ \alpha $ as the new value. To prevent over-fitting, regularization is applied so that the size of $ \alpha $ does not exceed $ 0.5 $.

\section{Results}

In this section, we present experimental results of applying the proposed method to synthetic data and to the binary classification task on Fashion MNIST \cite{xiao2017fashionmnist}. For the synthetic data, we generate samples as $ x \sim \mathcal{N}(0,I_{d}) $ and labels as $ y = \mathrm{sign}(\lVert x \rVert^{2} - \sqrt{d}) $, with $ d=10 $, split into 1000 training and 1000 test samples. For Fashion MNIST, we randomly selected 1000 images from each of the two specific classes and reduced their dimensions from 784 to 10 using principal component analysis (PCA). 500 RBMs with the same parameters were simultaneously embedded in subgraphs on D-Wave Advantage, as shown in Figure \ref{fig:boltzmann machine} to mitigate the cost of using a quantum annealer. In other words, we obtained two samples for each embedding in a total of 1000 samples. In this setup, the training time per iteration on the synthetic data was 0.4 $ s $; of that, the sampling time per iteration was only 483 $ \mu s $, and obtaining samples is known to be faster than with Markov chain Monte Carlo. \cite{Dixit2021}

\begin{figure}[H]
\centering
\includegraphics[width=\linewidth]{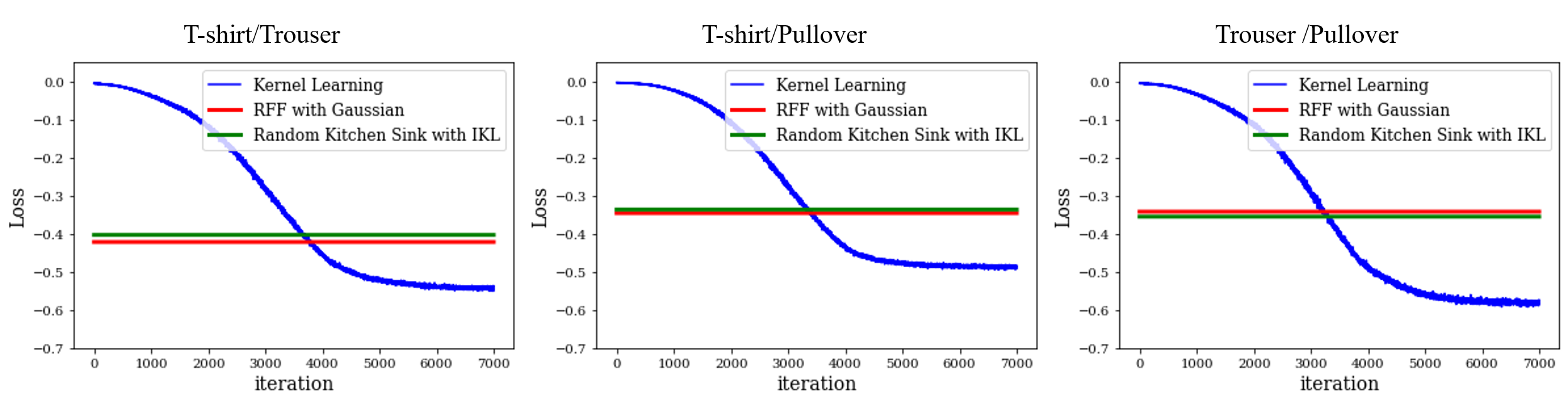}
\caption{(Color online) Loss function in kernel learning. 
The solid curve represents the value of the loss function obtained by kernel learning.
The horizontal reference lines represent the minimum loss obtained by RFF with a Gaussian spectral distribution and by RKS with IKL, respectively.
For the Gaussian RFF baseline, the mean and variance of the spectral distribution were optimized using Optuna. The reference lines are independent of the iteration number because they indicate optimized minimum loss values. Our method achieved a lower loss function value than both baseline methods for all pairs of classes. The three panels correspond to the classification tasks of T-shirt/Trouser, T-shirt/Pullover, and Trouser/Pullover, respectively.}
\label{fig:loss}
\end{figure}

\begin{figure}[H]
\centering
\includegraphics[width=\linewidth]{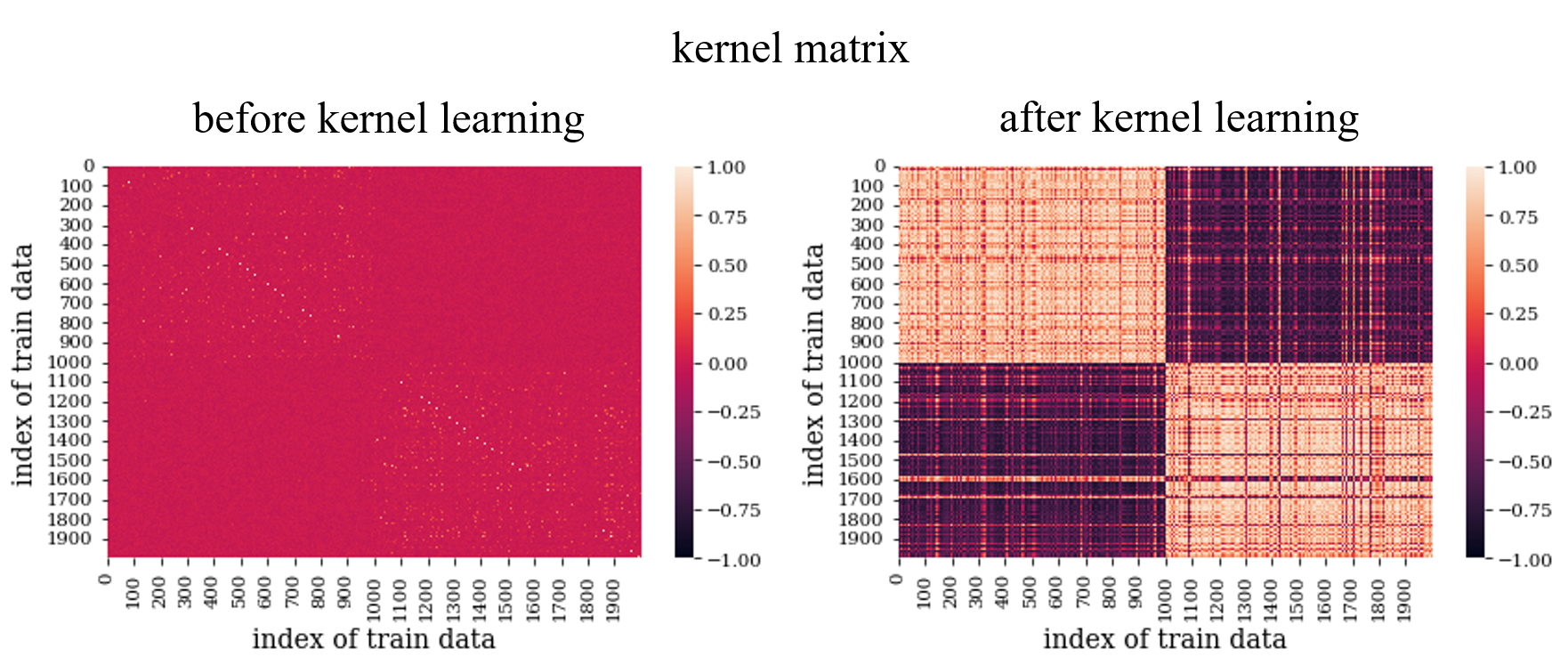}
\caption{(Color online) Kernel matrix before and after kernel learning. The left panel represents the kernel matrix before kernel training, and the right panel represents the kernel matrix after kernel training. 
The vertical and horizontal axes are the indices of the data, respectively. The data are arranged such that data with indices 0-999 have a label of $ +1 $ and data with indices 1000-1999 have a label $ -1 $. From equation (12), we can confirm that after learning, the value of kernel $ k(\bm{x}_{i},\bm{x}_{j}) $ for indices with the same label is larger and the value of kernel $ k(\bm{x}_{i},\bm{x}_{j}) $ for indices with different labels is smaller.}
\label{fig:kernel matrix}
\end{figure}

\begin{figure}[H]
\centering
\includegraphics[width=\linewidth]{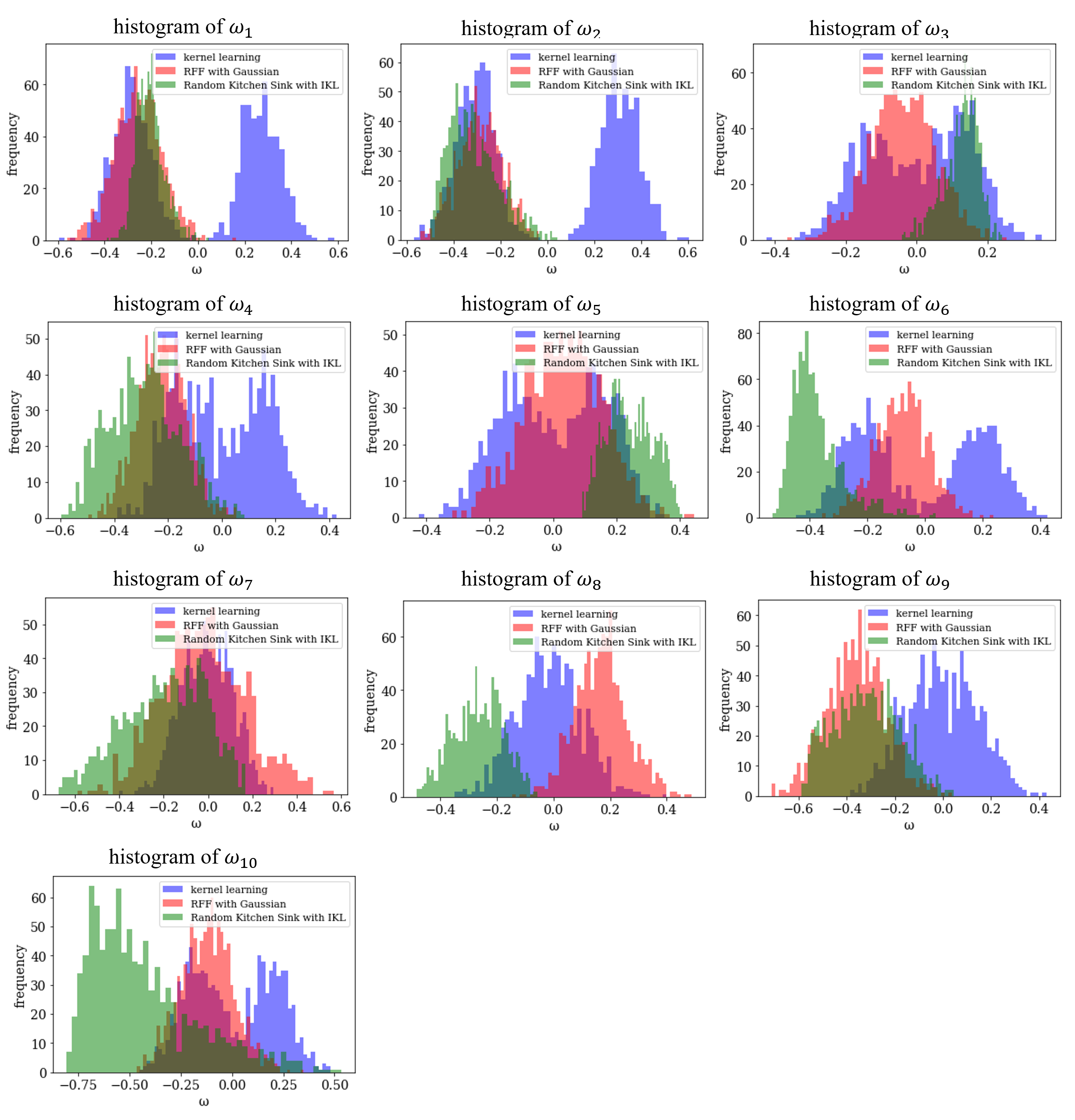}
\caption{(Color online) Histograms of all components of $ \bm{\omega} $ sampled from the trained GB-RBM, Gaussian RFF and RKS with IKL for the T-shirt/Trouser dataset. The Fashion MNIST image data are compressed to 10 dimensions using PCA, so $ \bm{\omega} $ is also 10-dimensional. Except for $ \omega_7 $, $ \omega_8 $ and $ \omega_9 $, the histograms from GB-RBM exhibit two peaks, clearly distinct from other methods.}
\label{fig:histogram}
\end{figure}

To evaluate the effectiveness of the proposed method, we conducted a comparative analysis using the following performance metrics:

\begin{itemize}
    \item Classification accuracy on the synthetic data (Table \ref{tab:syntheticdata_accuracy}): Comparison before and after kernel learning and RFF with Gaussian distribution;
    \item Classification accuracy (Table \ref{tab:mnistdata_accuracy}): Comparison of accuracy between our method before and after training, the RFF method using a Gaussian distribution and RKS with IKL \cite{Li2019}. For RKS with IKL, we use a multi-layer perceptron with input-noise 10 → hidden (4,4) → output 10, batch size 200, and the number of random features set to 200 (training) / 1000 (test).
    \item Loss function reduction (Figure \ref{fig:loss}): Demonstrates how our method effectively minimizes the loss function compared to RFF with Gaussian spectral distributions and RKS with IKL.
    \item Learned kernel matrix (Figure \ref{fig:kernel matrix}): Visual representation of the kernel matrix before and after training, showing the adaptation of the kernel to the data.
    \item Distribution of sampled frequencies (Figure \ref{fig:histogram}): Comparison between the spectral distribution learned by our method and the Gaussian-distributed RFF and RKS with IKL.
\end{itemize}

From Table \ref{tab:syntheticdata_accuracy}, the test accuracy on the synthetic data $ d=10 $ improves from $ 0.818 $ with Gaussian RFF to $ 0.846 $ after kernel learning, i.e., $ +0.028 $. Table \ref{tab:mnistdata_accuracy} shows that, for all three class pairs, the highest accuracy is achieved after kernel learning, outperforming both Gaussian RFF and RKS with IKL.
Table \ref{tab:mnistdata_accuracy} shows the prediction accuracy for each dataset. As can be seen from the table, the classification accuracy was highest after kernel learning.

\begin{table}
\centering
\begin{tabular}{|p{2.5cm}||c|c|c|c|}
\hline
{\hfill}Synthetic data{\hfill} & Before Kernel Learning & After Kernel Learning & RFF with Gaussian \\
\hline\hline
{\hfill}Train Data{\hfill} & 1.000 & 1.000 & 1.000 \\
\hline
{\hfill}Test Data{\hfill} & 0.677 & {\bf 0.846} & 0.818 \\
\hline
\end{tabular}
\caption{Synthetic data. Prediction accuracy before/after kernel learning and RFF with Gaussian. Test accuracy improves from 0.818 (RFF with Gaussian) to 0.846 (after), i.e., +0.028. Bold indicates the best.}
\label{tab:syntheticdata_accuracy}
\end{table}

\begin{table}
\centering
\begin{tabular}{|p{2.5cm}||c|c|c|c|}
\hline
{\hfill}T-shirt/Trouser{\hfill} & Before Kernel Learning & After Kernel Learning & RFF with Gaussian & RKS with IKL \\
\hline\hline
{\hfill}Train Data{\hfill} & 1.000 & 0.992 & 0.995 & 0.975 \\
\hline
{\hfill}Test Data{\hfill} & 0.929 & {\bf 0.990} & 0.988 & 0.975 \\
\hline
\end{tabular}

\begin{tabular}{|p{2.5cm}||c|c|c|c|}
\hline
{\hfill}T-shirt/Pullover{\hfill} & Before Kernel Learning & After Kernel Learning & RFF with Gaussian & RKS with IKL \\
\hline\hline
{\hfill}Train Data{\hfill} & 1.000 & 0.973 & 0.978 & 0.955 \\
\hline
{\hfill}Test Data{\hfill} & 0.926 & {\bf 0.961} & 0.959 & 0.949 \\
\hline
\end{tabular}

\begin{tabular}{|p{2.5cm}||c|c|c|c|}
\hline
{\hfill}Trouser/Pullover{\hfill} & Before Kernel Learning & After Kernel Learning & RFF with Gaussian & RKS with IKL \\
\hline\hline
{\hfill}Train Data{\hfill} & 1.000 & 0.995 & 0.997 & 0.972 \\
\hline
{\hfill}Test Data{\hfill} & 0.972 & {\bf 0.991} & 0.990 & 0.974\\
\hline
\end{tabular}
\caption{Fashion MNIST. Accuracy before/after kernel learning, Gaussian RFF and RKS with IKL. The bolded text indicates the highest prediction accuracy for each
dataset.}
\label{tab:mnistdata_accuracy}
\end{table}

Figure \ref{fig:loss} shows the loss function of kernel learning. As a reference, we also offer the value of the loss function of the RFF model with a Gaussian distribution and RKS with IKL for the training dataset. The RFF model with the parameters, namely the mean and variance, is optimized by Optuna\cite{Akiba2019} according to equation (12). The figure shows that kernel learning is more efficient in searching for the minimum loss function for all datasets.

Figure \ref{fig:kernel matrix} also shows the kernel matrices before and after kernel learning. It can be confirmed that kernel learning enables classification, as the kernel component $ k(\bm{x}_{i}, \bm{x}_{j} )$ is more significant for data with the same labels, and the kernel component $ k(\bm{x}_{i}, \bm{x}_{j}) $ is more minor for data with different labels.

Figure \ref{fig:histogram} shows the sampling of $ \bm{\omega} $ for our method, RFF with a Gaussian distribution, and RKS with IKL during training on the T-shirt/Trouser dataset.
It can be seen that when using the Gaussian distribution, the distribution of $ \bm{\omega} $ is Gaussian, whereas when using our method, a distribution with two peaks is generated. 
Therefore, it is possible to sample $ \bm{\omega} $ with a higher degree of freedom than when using a Gaussian distribution for the spectral distribution.

\section*{Discussion and Conclusion}
In this study, we formulated a method for kernel learning using GB-RBM. We showed that the quantum annealer can be used as a sampling machine for training the model, as part of our proposal for a novel algorithm that leverages quantum annealing. We applied the proposed method to the synthetic data and the Fashion MNIST dataset and demonstrated its feasibility for training using D-Wave’s quantum annealer. We confirmed that the prediction accuracy is comparable to that of the RFF model with a Gaussian kernel and RKS with IKL, but more importantly, the spectral distribution learned by our method exhibited a bimodal structure distinct from the other methods, as shown in Figure \ref{fig:histogram}. This highlights that our method can generate a data-adaptive kernel that captures distributional features inaccessible to fixed Gaussian RFF models. Furthermore, as shown in Table \ref{tab:mnistdata_accuracy} and Figures \ref{fig:loss} and \ref{fig:kernel matrix}, the improvements in prediction accuracy and the reduction in loss after training indicate that our method successfully adapts the kernel to the data.

Regarding dealing with kernel functions in quantum algorithms, related work includes fast computation of the Gaussian kernel\cite{Bishwas2020} and training RBF networks using the HHL algorithm\cite{Shao2019}. These studies propose an exponential speedup over classical algorithms, but they rely on a fault-tolerant quantum computer (FTQC), which is not achievable with today's quantum computers. On the other hand, our approach differs from those of these studies in that we tailor the kernel to fit the training data. While lacking mathematical guarantees for quantum speedup, our method is feasible on today's quantum annealers.

In the experiment, we conducted validation using a model size that can be easily executed on a quantum annealer; however, a future research topic is to clarify the relevance of the number of nodes in the hidden layer of the GB-RBM to the distribution of $\omega$, and the prediction accuracy needs to be investigated. 
In addition, we should clarify whether quantum annealers can achieve quantum speedup as sampling machines in our method. While a recent line of work, such as IKL, learns spectral distributions via adversarial training, we included RKS with IKL as a baseline and observed that our method consistently achieved the highest accuracy across all pairs (Table \ref{tab:mnistdata_accuracy}) and lower losses (Fig. \ref{fig:loss}). Future study includes exploring larger models and broader datasets for a more extensive comparison.

\begin{acknowledgment}

This work was financially supported by JSPS KAKENHI Grant No. 23H01432 and the programs for bridging the gap between R\&D and IDeal society (Society 5.0) and Generating Economic and social value (BRIDGE) and Cross-ministerial Strategic Innovation Promotion Program (SIP) from the Cabinet Office 23836436.

\end{acknowledgment}

\appendix
\section{The Gradient of Kernel Learning Loss Function}

In the section Kernel Learning with RBM, the gradient of the kernel learning loss function can be calculated as follows,
\begin{equation}
\frac{\partial L}{\partial \theta} = \frac{1}{N^{2}} \sum_{ij}{\frac{1}{S}\sum_{s}{y_{i} y_{j} e^{i\bm{\omega}_{s} (\bm{x}_{i} - \bm{x}_{j})}}} \left\{-\frac{\partial}{\partial \theta} E(\bm{\omega}, \bm{v}, \bm{h}| \theta) - \frac{1}{Z(\theta)} \frac{\partial}{\partial \theta} Z(\theta)\right\}_{\bm{\omega}_{s}}, \tag{A1}
\end{equation}
where, $ E(\bm{\omega}, \bm{v}, \bm{h} | \theta) $ is expressed as,
\begin{equation}
\label{energyfunc}
E(\bm{\omega}, \bm{v}, \bm{h} | \theta )= E(\bm{\omega}, \bm{v}, \bm{h} | \bm{a}, \bm{b}, \bm{c}, w, u, \bm{z}) = \sum_{i} \frac{(\omega_{i} - (a_{i} + \sum_{j}u_{ij}v_{j}))^{2}}{2\ e^{z_{i}}} - \sum_{j} {b_{j} v_{j}} - \sum_{k} {c_{k} h_{k}} - \sum_{jk} {v_{j} w_{jk} h_{k}}. \tag{A2}
\end{equation}
Each parameter is calculated using equation (\ref{energyfunc}) as follows,
\begin{equation}
\frac{\partial L}{\partial a_{i}} = \frac{1}{N^{2}} \sum_{ij}{\frac{1}{S}\sum_{s}{y_{i} y_{j} e^{i\bm{\omega}_{s} (\bm{x}_{i} - \bm{x}_{j})}}} \left\{\frac{\omega_{i} - (a_{i} + \sum_{j}u_{ij}v_{j})}{e^{z_{i}}} \right\}_{\bm{\omega}_{s}} \tag{A3}
\end{equation}
\begin{equation}
\frac{\partial L}{\partial b_{j}} = \frac{1}{N^{2}} \sum_{ij}{\frac{1}{S}\sum_{s}{y_{i} y_{j} e^{i\bm{\omega}_{s} (\bm{x}_{i} - \bm{x}_{j})}}} \left\{v_{j} - \frac{1}{Z(\theta)} \frac{\partial}{\partial b_{j}} Z(\theta)\right\}_{\bm{\omega}_{s}} \tag{A4}
\end{equation}
\begin{equation}
\frac{\partial L}{\partial c_{k}} = \frac{1}{N^{2}} \sum_{ij}{\frac{1}{S}\sum_{s}{y_{i} y_{j} e^{i\bm{\omega}_{s} (\bm{x}_{i} - \bm{x}_{j})}}} \left\{h_{k} - \frac{1}{Z(\theta)} \frac{\partial}{\partial c_{k}} Z(\theta)\right\}_{\bm{\omega}_{s}} \tag{A5}
\end{equation}
\begin{equation}
\frac{\partial L}{\partial u_{ij}} = \frac{1}{N^{2}} \sum_{ij}{\frac{1}{S}\sum_{s}{y_{i} y_{j} e^{i\bm{\omega}_{s} (\bm{x}_{i} - \bm{x}_{j})}}} \left\{\frac{\omega_{i}- (a_{i} + \sum_{j}u_{ij}v_{j})}{e^{z_{i}}} v_{j} \right\}_{\bm{\omega}_{s}} \tag{A6}
\end{equation}
\begin{equation}
\frac{\partial L}{\partial w_{jk}} = \frac{1}{N^{2}} \sum_{ij}{\frac{1}{S}\sum_{s}{y_{i} y_{j} e^{i\bm{\omega}_{s} (\bm{x}_{i} - \bm{x}_{j})}}} \left\{v_{j} h_{k} - \frac{1}{Z(\theta)} \frac{\partial}{\partial w_{jk}} Z(\theta)\right\}_{\bm{\omega}_{s}} \tag{A7}
\end{equation}
\begin{equation}
\frac{\partial L}{\partial z_{i}} = \frac{1}{N^{2}} \sum_{ij}{\frac{1}{S}\sum_{s}{y_{i} y_{j} e^{i\bm{\omega}_{s} (\bm{x}_{i} - \bm{x}_{j})}}} \left\{\frac{(\omega_{i} - (a_{i} + \sum_{j}u_{ij}v_{j}))^{2}}{2 e^{z_{i}}} - \frac{1}{2} \right\}_{\bm{\omega}_{s}}. \tag{A8}
\end{equation}
Here, $ \bm{v}, \bm{h} $ are sampled in the experiment as well as $ \bm{\omega} $. In other words, we obtain samples $ \left\{(\bm{\omega}_{1}, \bm{v}_{1}, \bm{h}_{1}), (\bm{\omega}_{2}, \bm{v}_{2}, \bm{h}_{2}), \cdots, (\bm{\omega}_{S}, \bm{v}_{S}, \bm{h}_{S})\right\} $. Therefore, when calculating each parameter, we use $ \bm{v} $ and $ \bm{h} $ associated with $ \bm{\omega} $.

\bibliographystyle{jpsj}
\bibliography{library}

\end{document}